\documentstyle[twocolumn,aps]{revtex}
\input{epsf.tex}

\def\be{\begin{equation}}
\def\ee{\end{equation}}
\def\ba{\begin{array}}
\def\ea{\end{array}}
\def\bea{\begin{eqnarray}}
\def\eea{\end{eqnarray}}
\begin{document}
\draft
\title{\bf Closed shell effects from the stability and instability of deformed and 
superdeformed nuclei against cluster decays in the mass regions 130-158 and 180-198}
\author{Raj K. Gupta, Sharda Dhaulta, Rajesh Kumar and M. Balasubramaniam} 
\address{Physics Department, Panjab University, Chandigarh-160014, India.}
\author{G. M\"unzenberg}
\address{Gesellschaft f\"ur Schwerionenforschung mbH, Planckstrasse 1, D-64291 Darmstadt, 
Germany.}
\author{Werner Scheid}
\address{Institut f\"ur Theoretische Physik, Justus-Liebig-Universit\"at,
Heinrich-Buff-Ring 16, D-35392 Giessen, Germany.}
\date{\today}
\maketitle
\begin{abstract}
The stability and/or instability of the deformed and superdeformed nuclei, $^{133-137}_{60}$Nd, 
$^{144-158}_{64}$Gd, $^{176-194}_{80}$Hg, and $^{192-198}_{82}$Pb parents, coming from three 
regions of different superdeformations, are studied with respect to the $\alpha$ and heavy 
cluster decays. The $\alpha$-decay studies also include the heavier $^{199-210}$Pb nuclei, for 
reasons of spherical magic shells at Z=82 and N=126. The calculations are made by using the 
preformed cluster-decay model, and the obtained $\alpha$-decay half-lives are compared with the 
available experimental data. Having met with a very good success for the comparisons of 
$\alpha$-decay half-lives and in giving the associated known magic or sub-magic closed shell 
structures of both the parent nuclei and daughter products, the interplay of closed shell 
effects in the cluster-decay calculations is investigated. The cluster-decay calculations also 
give the closed shell effects of known spherical magicities, both for the parent and daughter 
nuclei, and further predict new (deformed) closed shells at Z=72-74 and N=96-104 due to both 
the stability and instability of Hg and Pb parents against cluster decays. Specifically, a new 
deformed daughter radioactivity is predicted for various cluster decays of $^{186-190}$Hg and 
$^{194,195}$Pb parents with the best possible measurable cases identified as the $^8$Be and 
$^{12}$C decays of $^{176,177}$Hg and/or $^{192}$Pb parents. The predicted decay half-lives are 
within the measurable limits of the present experimental methods. The interesting point to note 
is that the parents with measurable cluster decay rates are normal deformed nuclei at the 
transition between normal and super-deformation.
\end{abstract}
\section{Introduction}
The $\alpha$-decay results have been used for identifying the shell closure effects for quite 
some time now, including even the very weak, sub-shell closures. For example, the Z=64 sub-shell 
was first noted by observing the systematics of $\alpha$-decay energies \cite{rasmuss53}, and 
later by a dip at Z=64 in the measured $\alpha$-decay reduced widths \cite{ott76}, of a few N=84 
isotones in its neighborhood. In the recent past, some of us and collaborators 
\cite{gupta91,gupta93,kumar94,kumar95,kumar96,gupta97,gupta99} have coupled the $\alpha$-decay 
studies with the exotic cluster-decay result of the observed spherical closed-shell daughter 
($^{208}$Pb or a neighbouring nucleus), called cluster radioactivity \cite{sandu80,rose84,gupta94}. 
This allowed us to predict two other new spherical closed-shell daughter radioactivies, namely 
$^{100}$Sn and $^{132}$Sn daughter radioactivies \cite{gupta93,kumar94,kumar95,kumar96}, and 
also a deformed daughter radioactivity at Z=74-76 and N=98-104 \cite{gupta97}. The spherical 
$^{100}$Sn daughter radioactivity has been emphasized also by Poenaru, Greiner and Gherghescu 
\cite{poenaru93}, and a couple of, so-far unsuccessful, experimental attempts have also 
been made to observe it as the ground-state decay of $^{114}$Ba nucleus produced in heavy-ion 
reactions \cite{oganess94,gugliel95}. This decay is now believed to belong to an excited 
compound nucleus decay, studied for $^{12}$C decay of $^{116}$Ba$^*$ \cite{commara00,gupta02}. 
Furthermore, the cluster decay studies are also used to point out the shell stabilizing effects 
of the parent nucleus \cite{gupta91,gupta97,gupta99}. Thus, both the cases of large and small 
decay rates (equivalently, the small and large decay half-lives) are found important, the large 
ones refering to closed shell effects of the daughter nucleus and the small ones to closed 
shell effects of the parent nucleus. In other words, taking a clue from the experiments, in a 
decay calculation, the presence of a known spherical or deformed daughter should result in a 
large decay rate (small decay half-life) or alternatively, a large decay rate (small decay 
half-life) should refer to the existence of a known or un-known (new), spherical or deformed, 
closed shell for the daughter nucleus. 

In the above mentioned calculations, we have so-far investigated the alpha and/or cluster 
decays of various neutron-deficient and neutron-rich rare-earths $_{54}$Xe to $_{64}$Gd 
\cite{gupta93,kumar94,kumar95,kumar96} and the even-A deformed and superdeformed 
$^{180-194}_{80}$Hg nuclei \cite{gupta97}. In mercury nuclei, the superdeformation begins at 
the $^{189}$Hg isotope, and the axes ratios are $\approx$1.7:1 \cite{firestone94}. Note that
the superdeformation here refers to the observation of (excited) superdeformed band(s) in 
these nuclei, though their ground-state deformations are not very much different from other 
neighbouring nuclei. This is illustrated in Fig. 1, where the data for ground-state quadrupole 
deformation parameter $\beta _2$ is taken from the calculations of M\"oller et al. 
\cite{moeller95}, since a similar data from experiments is not available for all the 
nuclei studied here. On the other hand, a deformed or normal deformed nucleus is one 
where superdeformed band(s) are not observed and it comes from the well known mass region 
$150<A<190$ of deformation. In the above stated nuclei, the closed shell effects of both 
the daughter products and the parent nuclei were analyzed. The stability of parent nuclei 
was also studied for the mass region A=68-82 \cite{gupta91,gupta99}, 
which includes several deformed and superdeformed nuclei (here, both in the ground states). 
However, there are several other regions of various deformations and superdeformations in the 
mass regions A=130-158 and 180-198 \cite{firestone94} whose decay characteristics still remain 
to be probed. The aim of this paper is to make a complete analysis of the decay properties of 
nuclei in these two mass regions, in order to get a general picture of how the deformed, in 
particular the superdeformed, nuclei behave against the $\alpha$ and heavier cluster decays. 
The superdeformed nuclei are expected to be more instable, though, like for the mass region 
A=68-82 \cite{gupta91,gupta99}, the following analysis does not seem to support this contention. 
Instead, the superdeformed nuclei are found to be rather poor $\alpha$ emitters, as compared to 
their lower mass, normal deformed nuclei. They are, however, shown to be the better $\alpha$ 
emitters than the heavier mass (heavier than superdefomed nuclei), normal deformed nuclei. The 
same is found true for cluster decay results. Such an un-expected situation is presented by the 
presence of known and/or un-known (new) closed shell effects of the daughter products. Also, the 
closed shell effects of either the protons or neutrons, as well as the neutron/proton asymmetry, 
of the parent nucleus play a role. 

The other nuclei that have superdeformations identical to those of $_{80}^{189-194}$Hg nuclei 
are the $^{191}_{79}$Au, $^{191-195}_{81}$Tl, $^{192-196,198}_{82}$Pb and $^{197}_{83}$Bi 
nuclei. Then, there are several rare-earths, from $_{62}$Sm to $_{66}$Dy and $_{68}$Er, which 
have even more strongly superdeformed shapes with axes ratios $\approx$2:1. Also, some other 
rare-earths, the $_{57}$La to $_{60}$Nd, have superdeformed species with axes ratios 
$\approx$1.5:1. In this paper, we choose to work specifically with both the odd- and even-A 
$^{133-137}_{60}$Nd, $^{144-158}_{64}$Gd and $^{182-198}_{82}$Pb parents, which comprise the 
three regions of different superdeformations mentioned above, along with some normal deformed 
nuclei. Note that $^{154-158}$Gd are known $\alpha$-stable nuclei, but are found to be of 
interest from the point of view of heavy-cluster instabilities and the associated closed shell 
effects (Section III.B.2). We have also included in our analysis here, the already studied 
\cite{gupta97} mercury nuclei, extended to both the odd- and even-A $^{176-194}_{80}$Hg, and 
the heavier $^{199-210}$Pb nuclei where some experimental data for $\alpha$-decays are 
available. Thus, the cases of both the normal deformed and superdefomed nuclei, and the 
spherical closed shell nuclei at and around Z=82, N=126, are covered in our study. Figure 1 
shows that all the superdeformed nuclei chosen here come from the transition (both lighter and 
heavier) regions of known deformed nuclei in the mass region $150<A<190$. 

The paper is organised as follows. The calculations are made by using the preformed 
cluster-decay model (PCM) of Gupta and collaborators \cite{gupta88,malik89,kumar97,gupta99a}
whose brief outline is presented in section II. Section III deals with the calculations and 
results obtained from this study. A summary of our results and conclusions are presented in 
section IV.

\section{The Preformed Cluster-decay Model} 
The preformed cluster-decay model (PCM) is a well established method for cluster decay studies.
We refer the reader to original papers \cite{gupta88,malik89,kumar97} or the reviews in Refs. 
\cite{gupta94,gupta99a} for complete details on the model. In the PCM, the decay constant 
$\lambda$ (or, inversely, the decay half-life time T$_{1/2}$) is the product of the cluster 
preformation probability P$_0$, the barrier impinging frequency $\nu_0$, and the barrier 
penetration probability P, 
\be 
\lambda={ln2\over {T_{1/2}}}=P_0{\nu_0}P.
\label{eq:1}
\ee
For calculating P$_0$ and P, the authors introduced, respectively, the dynamical collective 
coordinate of mass asymmetry $\eta$=(A$_1$-A$_2$)/A, with A=A$_1$+A$_2$, and relative separation 
R between the two fragments, via the stationary Schr\"odinger equation 
\be
H(\eta,R){\psi_n(\eta,R)}=E_n{\psi_n(\eta,R)}. 
\label{eq:2}
\ee
The potential part of the Hamiltonian in this equation is defined by 
\be 
V(\eta,R)=\sum_{i=1}^2 B_i(A_i,Z_i)+\frac{Z_1Z_2e^2}{R}+V_p, 
\label{eq:3}
\ee 
given as the sum of the experimental binding energies \cite{audi95} and the Coulomb and nuclear
proximity \cite{blocki77} potentials.  The fragmentation potential V($\eta$) and the scattering
potential V(R) are obtained from Eq. (\ref{eq:3}), respectively, for fixed R and $\eta$. The
R is fixed at the touching configuration, R=C$_t$=C$_1$+C$_2$, the C$_i$ being the S\"ussmann 
central radii $C_i=R_i-{1/{R_i}}$ (in fm) with R$_i$ as the equivalent spherical radii 
$R_i=1.28A_i^{1/3}-0.76+0.8A_i^{-1/3}$ fm; and $\eta$ is fixed by the emitted cluster. The 
charges Z$_i$ in (\ref{eq:3}) are fixed by minimizing the potential (without Vp) in the charge 
asymmetry coordinate $\eta_{Z}$=(Z$_1$-Z$_2$)/Z, with Z=Z$_1$+Z$_2$. 

In principle, the two coordinates are coupled, but in view of the defining equation (\ref{eq:1}), 
the Schr\"odinger equation (\ref{eq:2}) is solved in the decoupled approximation of $\eta$ and 
R-motions. Only the ground-state ($n$=0) solution is relevant for the cluster decay to occur in 
the ground-state of the daughter nucleus. Then, for $\eta$ motion, the properly normalized 
fractional cluster preformation probability is 
\be
P_0(A_2)={|\psi(\eta)|^2}{\sqrt{B_{\eta\eta}(\eta)}}\frac{2}{A},
\label{eq:4}
\ee
with $B_{\eta\eta}$ taken as the classical hydrodynamical mass of Kr\"oger and Scheid
\cite{kroeger80}. For the R-motion, we use the WKB approximation for calculating the 
penetrability P. In PCM, the penetration is considered to begin at $R=R_a=C_t$ and end at 
$V(R_b)$=Q-value of the decay.

Finally, the impinging frequency $\nu_0$ in the PCM is defined by considering that the total
kinetic energy, shared between the two fragments, is the positive Q-value. Then,
\be 
\nu_0={velocity\over {R_0}}=\frac{\sqrt{2Q/mA_2}}{R_0}.
\label{eq:5} 
\ee
Here R$_0$ is the equivalent spherical radius of the parent nucleus and mA$_2$ is the mass of
emitted cluster.

\section{Calculations}
In this section, we present our calculations first for $\alpha$-decay, compared with the
experimental data, wherever available. Then, we analyze the cluster-decay calculations with a 
view to look for the role of known magic shells in both the daughter and parent nuclei, and the 
possible new closed-shell daughter products presenting the signatures of a new radioactivity, 
if any. The cluster decay calculations are presented separately for each set of nuclei.

Figures 2 and 3 show the fragmentation potentials for Nd and Pb nuclei, as the representatives 
of the two mass regions (A=130-158 and 180-198) studied here. The experimental binding energies 
used are from the 1995 tables of Audi and Wapstra \cite{audi95}. We notice that in each case, 
the potential energy minima occur at $^4$He and other N=Z, $\alpha$ nuclei, as well as at 
N$\ne$Z, non-$\alpha$ nuclei for all the parents in the heavier mass region (A=180-198) and for 
only the heavier parents in the lighter mass region A=130-158. This means that the 
$\alpha$-nuclei decay products are energetically more favoured for the lighter isotopes of Nd 
and Gd parents in the lighter mass region A=130-158, and the non-$\alpha$ decay products become 
equally favourable for all the Hg and Pb parents in the heavier mass region A=180-198 and for 
the heavier isotopes of Nd and Gd parents in A=130-158 mass region. We are interested only in 
the potential energy minima because the preformation factors P$_0$ for nuclei at the minima are 
the largest, compared to their neighbours, as is depicted in Fig. 4 for Nd and Gd and in Fig. 5 
for Hg and Pb parents, where, for some clusters belonging to the minima in the fragmentation 
potentials, the (negative) logarithm of the preformation probability P$_0$ is plotted as a 
function of the mass number of the parents. We notice that in all cases, like in our earlier 
calculations \cite{gupta91,gupta93,kumar94,kumar95,kumar96,gupta97,gupta99}, the preformation 
factor is largest for $^4$He and it goes on decreasing as the size of the cluster increases.  
Another point of interest to note in these figures is the change of clusters for the heavier 
parents (see the dashed parts of lines). For example, in both the Figs. 4 and 5, the cluster 
$^{16}$O for $^{144-154}$Gd and $^{176-193}$Hg nuclei changes to $^{16}$C for heavier 
$^{155-158}$Gd and $^{194}$Hg nuclei. However, then the Q-value (for $^{16}$C cluster 
combination) is so small (see Fig. 6) that the penetrability is almost negligible (and hence of 
not much interest to include such clusters any further in our analysis). Figure 6 also reveals
that the Q-value is negative (or nearly zero) for $\alpha$-decay of $^{146}$Gd, and for both
the $\alpha$ and $^8$Be decays of Gd nuclei heavier than $^{154}$Gd. The fact that the 
penetrabilities P are small ($-log_{10}P$ large) for smaller Q-values, is evident from Figs. 7 
and 8, which give, similar to Figs. 4 and 5, the results of our calculation for the barrier 
penetrability P. We further notice in Fig. 7 that the penetrability P is in general small 
(large $-log_{10}P$) for non-$\alpha$ clusters in the light mass region A=130-158. 
The combined effect of the preformation probability P$_0$ and penetrability P gives the 
measurable decay half-life time T$_{1/2}$, since the impinging frequency $\nu_0$ is almost 
constant. The resulting T$_{1/2}$ are presented in Figs. 9 and 10, where their logarithms are 
plotted with respect to the mass number of the parent nuclei. The structural information 
obtained from these calculations for each set of parents is discussed separately in the 
following sub-sections. Note, however, in Fig. 9(b) that $^{154}$Gd is almost stable against
$\alpha$ and $^8$Be decays (large T$_{1/2}$-values), but could be of interest for other heavier 
cluster decays, as is discussed in section III.B.2.

\subsection{The $\alpha$-decay results}
We have noted above that the preformation factor P$_0$ is largest for $^4$He. This is of the
order of $10^{-5}-10^{-8}$ for all superdeformed $^{133-137}$Nd, superdeformed and some heavier
mass, normal deformed $^{144-154}$Gd nuclei (superdeformation in Gd nuclei stops at $^{150}$Gd, 
and Gd nuclei beyond $^{154}$Gd are $\alpha$-stable, $Q_{\alpha}<$0; Q-value is small but
positive for $\alpha$-decay of $^{154}$Gd, though experimentally it is a known $\alpha$-stable
nucleus), all superdeformed and some lighter mass (mainly odd-A), normal
deformed $^{183-194}$Hg (here superdeformation begins at $^{189}$Hg) and all
superdeformed $^{192-198}$Pb nuclei. However, P$_0$ is much larger, $\sim 10^{-3}$, for almost 
all lighter mass, normal deformed $^{176-188}$Hg and $^{182-191}$Pb nuclei. This suggests that 
superdeformed nuclei are the poorer $\alpha$-emitters, as compared to their light mass, normal 
deformed species. Interesting enough, the same result is born out in the calculated 
$\alpha$-decay half-lives T$_{1/2}^{\alpha}$, plotted in Fig. 11 as 
$log_{10}$T$_{1/2}^{\alpha}$ versus parent mass number, and compared with the 
available experimental data (taken from Refs. \cite{royer00,garcia00}). We have also included 
here (see inset, Fig. 11) the other heavier isotopes of Pb in order to include the lone 
experimetal data for $^{210}$Pb, in the neighborhood of doubly magic $^{208}$Pb nucleus. The 
Fig. 11 presents not only the interesting comparison of calculated results with experiments, 
but also interesting shell structure effects of both the parent nuclei and their daughter 
products which are mostly known but not yet observed experimentally via $\alpha$-decay studies 
(the $\alpha$-decay experimental data are not yet complete). We have also compared our results
and the available experimental data with another calculation due to the generalized liquid drop 
model (GLDM) \cite{royer00} in Table 1. We notice that the PCM and the GLDM calculations give 
identical results, both within one order of magnitude of the experimental data.
 
The following results are evident from Fig. 11: (i) The PCM calculations compare nicely with 
experiments, showing even the (small) odd-even effects in both the light mass, normal deformed 
$^{176-188}$Hg and $^{182-191}$Pb nuclei. The odd-even effects seem to become weaker in all the 
superdeformed and heavier normal deformed nuclei studied here (see e.g. $^{148-152}$Gd or
$^{189-194}$Hg nuclei). The data are not yet enough to show the odd-even effects as clearly as 
are given by the calculations. The comparison between the experimental data and our calculations
for all the five $^{148-152}$Gd isotopes (containing no apparent odd-even effects) and the 
single $^{210}$Pb parent in very heavy mass region are also particularly striking. (ii) The 
light mass, normal deformed $^{176-188}$Hg and $^{182-191}$Pb parents are clearly the better 
$\alpha$ emitters (smaller $T_{1/2}^{\alpha}$ values), as compared to their heavier and 
superdeformed counterparts and the superdeformed $^{133-137}$Nd and $^{144-154}$Gd parents. 
This also includes the heavier, normal deformed $^{151-154}$Gd nuclei. However, the
superdeformed nuclei are better $\alpha$ emitters than the heavier mass, normal deformed
nuclei. In other words, the superdeformed
nuclei are though better $\alpha$ emitters than the heavier mass, normal deformed (heavier 
than the superdeformed) nuclei, but both are poorer $\alpha$ emitters as compared to the 
light mass, normal deformed nuclei. (iii) The large (peaking of) $T_{1/2}^{\alpha}$-values 
for $^{146}$Gd and $^{207,208}$Pb parents show the shell closure effects (strong stability) 
of magic N=82 and 126 coupled with their semi-magic Z=64 and magic Z=82, respectively. 
(iv) The above noted peaking effect of $T_{1/2}^{\alpha}$ at the spherical magic 
shells of the parents could be interpreted as the signature of the change of a shell from the 
deformed to the spherical one or one can say the presence of a (spherical or deformed) 
closed-shell daughter at the bottom(s) or valley(s) of such a peak. One such result is evident 
for the minimum (valley) at $^{148}$Gd, which refers clearly to N=82 spherical daughter 
$^{144}_{62}$Sm for its $\alpha$ decay. A similar minimum at $^{134}$Nd could be due to the 
mid-shell effect of magic Z=50 and sub-magic Z=64 for the (weakly deformed) daughter 
$^{130}_{58}$Ce. Such peaking effects are also visible in the superdeformed Hg and Pb nuclei. 
Apparently, all these results clearly establish the credentials of PCM for its possible 
predictions for the heavier-cluster decays and the associated shell structure effects, studied 
in the following sub-section.

\subsection{The heavy cluster-decays and closed-shell effects}
We have seen above, and is also known from our earlier calculations 
\cite{gupta93,kumar94,kumar95,kumar96,gupta99}, that as the N:Z ratio of parent nuclei 
increases, the N$\ne$Z, non-$\alpha$ nuclei cluster emissions become equally, or even more, 
probable as compared to N=Z, $\alpha$ nuclei cluster emissions (see, e.g., the crossing over of
the curves for $^{12}$C and $^{14}$C clusters in Fig. 4(b) for heavier Gd parents or in Fig. 
5(a) for Hg parents; P$_0$ becomes larger for $^{14}$C, as compared to that for $^{12}$C).
Hence, the $\alpha$-nucleus cluster emission effects must be more prevalent for parents in the 
low mass region A=130-158 and the non-$\alpha$ nuclei cluster emissions for parents in the 
heavier mass region A=180-198. We know that all the radioactive exotic cluster-decays from the 
parent masses A$>$222 consist of only non-$\alpha$ nuclei clusters \cite{gupta94}, such as 
$^{14}$C, $^{18,20}$O, $^{23}$F, $^{22,24,26}$Ne, $^{28,30}$Mg and $^{32,34}$Si. Secondly, it is 
interesting to note that, though the cluster preformation factors P$_0$ are of similar orders 
for both the chosen regions of parent nuclei (compare Figs. 4 and 5), the penetrabilities P are 
much smaller (larger $-log_{10}$P) for parents in the lighter mass region A=130-158 (compare 
Figs. 7 and 8). This means that, like for $\alpha$-decays, the cluster-decay rates for parents 
in the lighter mass region are also expected to be smaller (larger cluster-decay half-lives) 
than for parents in the heavier mass region. In other words, the parents in lighter mass region 
A=130-158 are likely to be more stable against cluster decays, than the parents in heavier mass 
region A=180-198. We discuss these results in the following for each set of parent nuclei 
separately. 

\subsubsection{Nd parents}
First of all we look at the calculations in Figs. 4(a) and 7(a), respectively, for the 
preformation probability P$_0$ and penetrability P for Nd parents. We notice that -log$_{10}$P$_0$ 
for the lightest two clusters $^8$Be and $^{12}$C are structure-less (remains almost constant), 
but for heavier clusters develop into maxima (minima for P$_0$) each at the even-A parents 
$^{134}$Nd and $^{136}$Nd which grow as the size of the cluster increases. On the other hand, 
for the penetrability, -log$_{10}$P is structure-less for all clusters, except for a steep rise 
for $^8$Be decay and small maximum (minimum for P) at $^{134}$Nd (or a minimum at $^{135}$Nd) 
for its $^{14}$C decay. The fact that these maxima and minima are simply the result of an 
odd-even effect in Nd nuclei is evident from the almost constant cluster decay half-lives 
T$_{1/2}^{c}$ in Fig. 9(a). The notable exception is again for $^8$Be decay, where the decay 
half-life is an ever increasing function of parent mass, with T$_{1/2}^{c}>$10$^{100}$(s) and
hence stable against such a decay. The interesting result is that some heavy clusters, like 
$^{30,32}$Si, have decay half-lives of the same order as for the light clusters like $^{12}$C 
and $^{16}$O. These are apparently due to, say, the neighbouring Z=50 magic shell in 
$^{119}_{52}$Te daughter, or the mid-shell effects of the known neutron magic shells, in 
$^{16}$O decay of $^{135}$Nd parent. The decay half-lives are, however, large 
$\sim$10$^{50}$ (s). In other words, the Nd parents are as stable, rather more stable, against 
cluster decays as they are against $\alpha$-decays.

\subsubsection{Gd parents}
For Gd parents, the P$_0$ in Fig. 4(b) seem to behave smoothly, except for a small minimum at 
$^{148}$Gd parent, which turn into a minimum at $^{145}$Gd and a maximum at $^{148}$Gd parent 
for the heavier clusters. Also, the division between the superdefomed and normal deformed nuclei 
is evident at $^{152}$Gd where P$_0$ increases suddenly ($-log_{10}$P$_0$ decreases) for all the 
clusters and stays nearly independent of the mass of normal deformed parents (except for small
oscillations, the odd-even effect). On the other hand, the P in Fig. 7(b) show the maxima, 
minima structure for different light clusters at any one of the parent nuclei $^{146-152}$Gd. 
The heavier clusters are all peaked around $^{146}$Gd. These results combine to give four 
significant minima for T$_{1/2}^{c}$ in Fig. 9(b): one at $^{150}$Gd for $^8$Be decay, another 
at $^{152}$Gd for $^{12}$C decay, the third one at $^{154}$Gd for $^{14}$C decay and finally the 
fourth one at $^{156}$Gd for $^{18}$O decay. All these minima refer to N=82 magicity of the 
respective daughters $^{142}_{60}$Nd, $^{140}_{58}$Ce, $^{140}_{58}$Ce and $^{138}_{56}$Ba. In 
other words, these are the only four isotopes of Gd ($^{152-156}$Gd) which are prone to heavier
cluster decays, though the predicted decay half-lives are large $\sim$10$^{43}$-10$^{67}$(s). 
Note that one of these parents ($^{150}$Gd) is a superdeformed nucleus whereas the other three 
heavier ones ($^{152,154,156}$Gd) are normal deformed nuclei. Also, of these four, $^8$Be and 
$^{12}$C decays of $^{150}$Gd and $^{152}$Gd, recpectively, are more probable (smaller 
T$_{1/2}^{c}$). Then, there are some maxima appearing in Fig. 9(b), mainly at $^{146}$Gd, 
which refer to the stability of this parent nucleus against the cluster decays due to its N=82 
shell closure. Note that $^{146}$Gd is already stable against $^8$Be decay ($Q<0$). Thus, the 
structure effects of both the parent(s) and daughters come into play in the cluster-decay 
properties of Gd nuclei, but the predicted cluster decay half-lives are beyond the limits of 
the present measurements, which go only upto $\sim$10$^{28}$(s) \cite{gupta94}.

\subsubsection{Hg parents}
In this sub-section, we discuss the results of our calculations for normal deformed 
$^{176-188}$Hg and superdeformed $^{189-194}$Hg nuclei, presented in Figs. 5(a), 8(a) and 10(a)
for P$_0$, P and T$_{1/2}^{c}$, respectively. First of all, we notice a number of minima and 
maxima in Fig. 5(a) for P$_0$, which correspond to the odd-even effects of the parents. Then, 
the P$_0$ increases suddenly ($-log_{10}$P$_0$ decreases) near the transition point of deformed 
to superdeformed region where it has the largest value for almost all the cluster preformations 
in $^{188}$Hg, the normal deformed nucleus at the transition. As we shall see below for the 
T$_{1/2}^{c}$ calculations, this result corresponds exactly to the one observed for the 
$\alpha$-decay half-lives of Gd isotopes in Fig. 11, i.e. of the change of shape in going from a 
maximum (peaking) to the minimum (valley). On the other hand, the P are nearly smooth functions 
of the parent mass, except for a noticeable minimum (enhanced penetrability) at $^{185}$Hg, 
and/or for some clusters at $^{189}$Hg, in Fig. 8(a). The resulting T$_{1/2}^{c}$ in Fig. 10(a) 
show interesting maxima and minima, like for P$_0$, for normal deformed $^{176-185}$Hg nuclei, 
referring to the larger stability of the even parents (at maxima) relative to the odd parents 
(at minima). Then, a (broad) minimum or valley of instability (smaller values next to a maximum 
in T$_{1/2}^{c}$) occurs for the normal deformed and superdeformed transitional nuclei 
$^{186-190}$Hg. 

The above noted stability of even-A $^{176-185}$Hg nuclei point to the closed shell effects of 
these parents at Z=80 (in the neighborhood of magic Z=82) coupled with a magic or semi-magic
nature of their neutron shells with N=96,98,100,102 and 104. On the other hand, the instabilty 
of $^{186-190}$Hg parents, against various clusters, reflect the closed shell effects of the 
daughter nuclei, like $^{178-182}_{76}$Os, $^{174-178}_{74}$W, .., etc., referring to Z=76,74,.. 
and N=106,104,102,.. closed or near-closed shells. Coupling the results of both the stability 
and instability in this region, with the fact that $^8$Be and $^{12}$C are shown as the most 
probable cluster decays (smallest T$_{1/2}^{c}$-values), the Z=76 or 74 and N=96-104 seem 
to point to the major (deformed) closed shells. The same result was obtained in our earlier 
studies \cite{gupta93,gupta97} and supports the structure calculations of other authors 
\cite{hannachi88,dracoulis88}. Also, Fig. 10(a) shows that the best measurable $^8$Be and 
$^{12}$C decays come from $^{176,177}$Hg, with T$_{1/2}^{^8Be}\sim$10$^{18}$(s) and 
T$_{1/2}^{^{12}C}\sim$10$^{23}$(s), which are well within the limits of present day 
experiments.      
 
\subsubsection{Pb parents}
The calculations for Pb parents are presented in Figs. 5(b), 8(b) and 10(b), respectively, for
P$_0$, P and T$_{1/2}^{c}$. The P$_0$ are almost constant, except for a small enhancement 
(minimum $-log_10$P$_0$) in the case of $^{196}$Pb parent for all cluster configurations. The 
same is true for the P, except that the enhancement is now for the $^{195}$Pb parent. The net 
result is a shallow minimum in T$_{1/2}^{c}$ for heavier clusters, like $^{20}$O and $^{24}$Ne,
from $^{194}$Pb or $^{195}$Pb parents, which means the shell stabilizing effects of the daughters 
$^{174,175}_{74}$W and $^{170,171}_{72}$Hf. This means supplementing the above noted results for 
Hg isotopes that the major closed shells could even occur at Z=72,74, N=98-100. The minimum 
decay half-lives are for the $^8$Be and $^{12}$C decays of $^{192}$Pb, both $\sim$10$^{37}$, 
which are still not within the reach of experiments. 

\section{Summary of Results and conclusion}
We have made a systematic study of the $\alpha$- and heavy-cluster decays of nuclei with masses
A=130-158 and 180-198, comprising three regions of superdeformations to different orders. 
Specifically, the $^{133-137}$Nd, $^{144-158}$Gd, $^{176-194}$Hg and $^{192-198}$Pb nuclei are 
studied, which also include the normal deformed nuclei on both the lighter and heavier sides of
the superdeformed nuclei. Furthermore, for $\alpha$-decay of Pb nuclei, we have also considered 
the very heavy isotopes upto $^{210}$Pb which means including also the spherical closed shell 
effects of $^{208}$Pb with doubly magic Z=82 and N=126. The main idea of this work  is to look 
for new (spherical or deformed) closed shells via cluster decay studies or, in other words, 
the possible signatures of any new cluster radioactivity in the superdeformed nuclei. A parent 
nucleus is stable against $\alpha$ and other cluster-decays, if the decay half-lives are large. 
On the other hand if the decay half-life is small (measurable or close to measurable) then, 
in view of the so-far observed radioactive decays, it must refer to the closed shell effects 
of the daughter product(s). Based on such an analysis for both the $\alpha$ and heavier cluster 
decays, we have obtained the following results from the chosen nuclear mass regions.

The superdeformed nuclei are better $\alpha$ emitters, as compared to their heavier mass, normal 
deformed isotopes, though both of these are poorer $\alpha$ emitters than their lighter mass, 
normal deformed species. The closed shell effects of the parents (for $^{146}$Gd and $^{208}$Pb
parents), at the spherical sub-magic Z=64, magic N=82 and doubly magic Z=82, N=126, are evident
in terms of the maxima or peaking of the $\alpha$-decay half-lives, whereas the same for 
daughter nuclei $^{130}$Ce and $^{144}$Sm (for $^{134}$Nd and $^{148}$Gd parents, respectively) 
are given as minima or valleys in $\alpha$-decay half-lives due to the mid-shell effect of the 
magic Z=50 and sub-magic Z=64, and the N=82 magic shell. These are the known shell closure 
effects, which apparently are nicely reproduced in the PCM calculations.

The calculated cluster-decay half-lives show that the lighter mass region A=130-158 is more 
stable (larger T$_{1/2}^{c}$ values) than the heavier mass region A=180-198. The light mass
region presents the closed shell effects of known magic Z=50 or N=82 shells (for $^{16}$O decay 
of $^{135}$Nd, $^8$Be decay of $^{150}$Gd, $^{12,14}$C decays of $^{152,154}$Gd, respectively, 
and $^{18}$O decay of $^{156}$Gd), with the predicted cluster decay half-lives beyond the 
present measurable limits of the experiments(T$_{1/2}^{c}\sim$10$^{43}$(s) or more). On the 
other hand, the heavier mass region A=180-198 present not only the stability effects of 
neighbouring Z=82 magic shell for even-A $^{176-185}$Hg parents, but also interesting new 
possibilities of deformed-daughter cluster radioactivity at Z=72-76 and N=96-104 for 
$^{186-190}$Hg and $^{194,195}$Pb. In other words, new deformed magic shells are likely to occur 
at Z=72-76 and N=96-104 for various cluster decays of $^{186-190}$Hg and $^{194,195}$Pb. The 
best, observable cases are predicted to be $^8$Be and $^{12}$C decays of $^{176,177}$Hg and/or 
$^{192}$Pb parents (with T$_{1/2}^{c}\sim$10$^{18}$ and 10$^{23}$ (s), respectively). 
The interesting point is that in both the nuclear regions under study, more of these nuclei are 
normal deformed nuclei rather than superdeformed ones. In fact, they lie at the 
transition between the normal deformation and superdeformation, but more towards the region of 
normal deformation. The best possible cases also come from the normal deformed regions. In other 
words, the region of the change of shape, i.e. the valley(s) in the immediate neighbourhood of 
a peak, seems to be the criterion for the location of new magicities or for the new cluster 
radioactivity(ies). 

\section{ACKNOWLEDGMENTS}
This work is supported in parts by the Council of Scientific and Industrial Research (CSIR), 
India, and the VW-Stiftung, Germany.


\newpage
\par\noindent {\bf Figure Captions:}
\vskip 6pt
\begin{description}
\item {Fig. 1.} The variation of ground-state quadrupole deformation parameter $\beta _2$ with
mass number for the selected Nd, Gd, Hg and Pb parent nuclei. The data are from the 
calculations of M\"oller et. al \cite{moeller95}. The region of nuclei where superdeformed bands 
are observed is marked in each case.
\item {Fig. 2.} The mass fragmentation potentials as a function of the mass number of light
fragments, for the superdeformed isotopes of Nd parents. The calculations are made at the 
touching configuration R=C$_1$+C$_2$=C$_t$ and by using experimental binding energies 
\cite{audi95}. Only the light fragments (clusters) at minima are marked. 
\item {Fig. 3.} The same as for Fig. 2, but for both the normal deformed and superdeformed 
isotopes of Pb parents. 
\item {Fig. 4.} The logarithms of the cluster preformation probability P$_0$ as a function of 
the mass number of (a) Nd, and (b) Gd parents, for different clusters. For the same cluster 
mass number, the dashed line shows P$_0$ if the charge number is changed.
\item {Fig. 5.} The same as for Fig. 4, but for (a) Hg, and (b) Pb parents.
\item {Fig. 6.} The variation of Q-value with mass number for (a) Gd and (b) Hg parents. The 
binding enegies used are the experimental binding energies \cite{audi95}. 
\item {Fig. 7.} The same as for Fig. 4, but for the penetrability P.
\item {Fig. 8.} The same as for Fig. 7, but for (a) Hg, and (b) Pb parents.
\item {Fig. 9.} The same as for Fig. 4, but for the logarithms of cluster decay half-life,
$log_{10}$T$_{1/2}^{c}$(s).
\item {Fig. 10.} The same as for Fig. 9, but for (a) Hg, and (b) Pb parents.
\item {Fig. 11.} The logarithm of the calculated $\alpha$-decay half-lives, 
$log_{10}$T$_{1/2}^{\alpha}$(s), as a function of the parent mass number for various isotopes of 
Nd, Gd, Hg and Pb nuclei, compared with the experimental data (taken from Refs. 
\cite{royer00,garcia00}). The inset shows the same calculation for the heavier isotopes of Pb, 
where the experimetal data for only $^{210}$Pb is known.
\end {description}
\newpage
{\footnotesize
\begin{table}
{\bf Table 1:}
{The logarithms of $\alpha$-decay half-lives and other characteristic quantities
calculated by using the preformed cluster model (PCM) of Gupta and
collaborators. The impinging frequency $\nu_0$ is nearly constant, of the order of 
${10}^{21} s^{-1}$. The PCM calculations are compared with the available GLDM 
calculations of Royer \cite{royer00} and the experimental data.
}
\begin{center}
\begin{tabular}{|c|c|c|c|c|c|c|}\cr
Parent&Q-value &\multicolumn{2}{c|}{PCM} & \multicolumn{3}{c|}{$log_{10} T_{1/2} (s)$} \\ \hline
&(MeV)&$P_0$ & P &PCM&GLDM&Expt.\\ \hline
$^{148}Gd$ & 3.27  & $3.10\times 10^{-05}$ & $1.38\times 10^{-28 }$  & $10.90 $ &9.68&9.36\\
$^{149}Gd$ & 3.10  & $2.84\times 10^{-07}$ & $2.84\times 10^{-29 }$  & $13.64 $&11.15&11.21\\
$^{150}Gd$ & 2.81  & $4.13\times 10^{-07}$ & $7.13\times 10^{-31 }$ & $15.10 $&14.09&13.75\\
$^{151}Gd$ & 2.65  & $9.41\times 10^{-08}$ & $5.92\times 10^{-32 }$& $16.83 $ &15.92&15.11\\
$^{152}Gd$ & 2.21  & $1.02\times 10^{-05}$ & $2.78\times 10^{-36 }$& $19.17$ &22.12&21.54\\
$^{176}Hg$ & 6.93  & $6.21\times 10^{-03}$ & $9.63\times 10^{-18 }$& $-2.38$ &-1.76&-1.7\\
$^{177}Hg$ & 6.74  & $1.36\times 10^{-04}$ & $4.69\times 10^{-18 }$& $-0.40$ &-1.17&-0.77\\
$^{178}Hg$ & 6.58  & $1.82\times 10^{-03}$ & $2.39\times 10^{-18 }$& $-1.23$ &-0.60&-0.44\\
$^{179}Hg$ & 6.43  & $9.57\times 10^{-05}$ & $1.22\times 10^{-18 }$& $0.35 $ &-0.04&0.32\\
$^{181}Hg$  & 6.29  & $5.36\times 10^{-05}$ & $6.04\times 10^{-19 }$ & $0.91 $ &0.49&1.32\\
$^{182}Hg$ & 6.00  & $1.29\times 10^{-03}$ & $1.23\times 10^{-19 }$  & $0.23 $ &1.68&1.85\\
$^{183}Hg$ & 6.04  & $1.83\times 10^{-06}$ & $1.62\times 10^{-19 }$& $2.96 $ &1.50&1.57\\
$^{184}Hg$ & 5.66  & $6.60\times 10^{-04}$ & $1.49\times 10^{-20 }$& $1.45 $ &3.20&3.37\\
$^{186}Hg$ & 5.21  & $4.02\times 10^{-05}$ & $4.49\times 10^{-22 }$& $4.21 $ &5.40&5.73\\  
$^{182}Pb$ & 7.08  & $6.50\times 10^{-03}$ & $1.57\times 10^{-17 }$  & $-2.61 $ &-1.59&-1.26\\
$^{189}Pb$ & 5.86  & $1.12\times 10^{-07}$ & $2.57\times 10^{-20 }$& $4.98 $ &3.17&4.11\\
$^{191}Pb$ & 5.41  & $4.64\times 10^{-07}$ & $8.29\times 10^{-22 }$& $5.87 $ &5.31&5.78\\
$^{210}Pb$ & 3.79  & $2.48\times 10^{-09}$ & $2.63\times 10^{-30 }$& $16.74 $ &16.00&16.57\\
\cr
\end{tabular}  
\end{center}
\end{table}
}

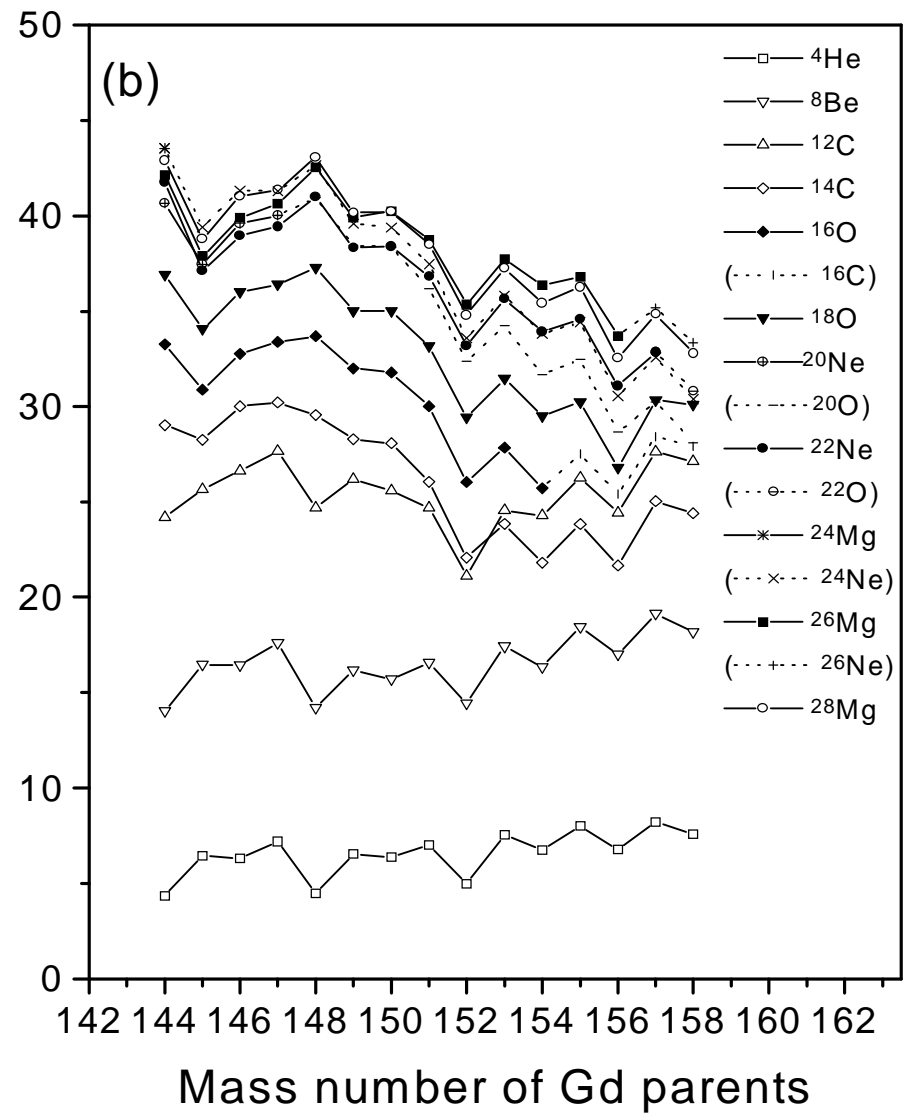
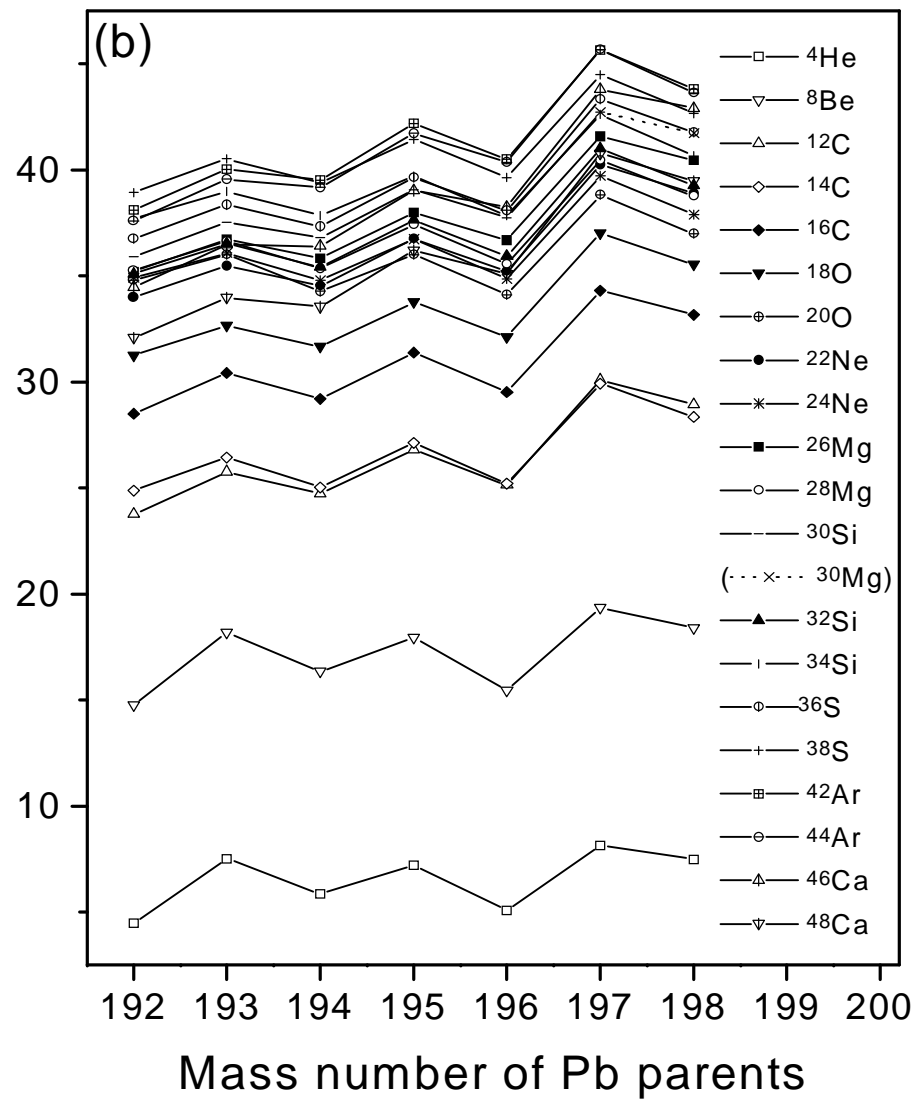
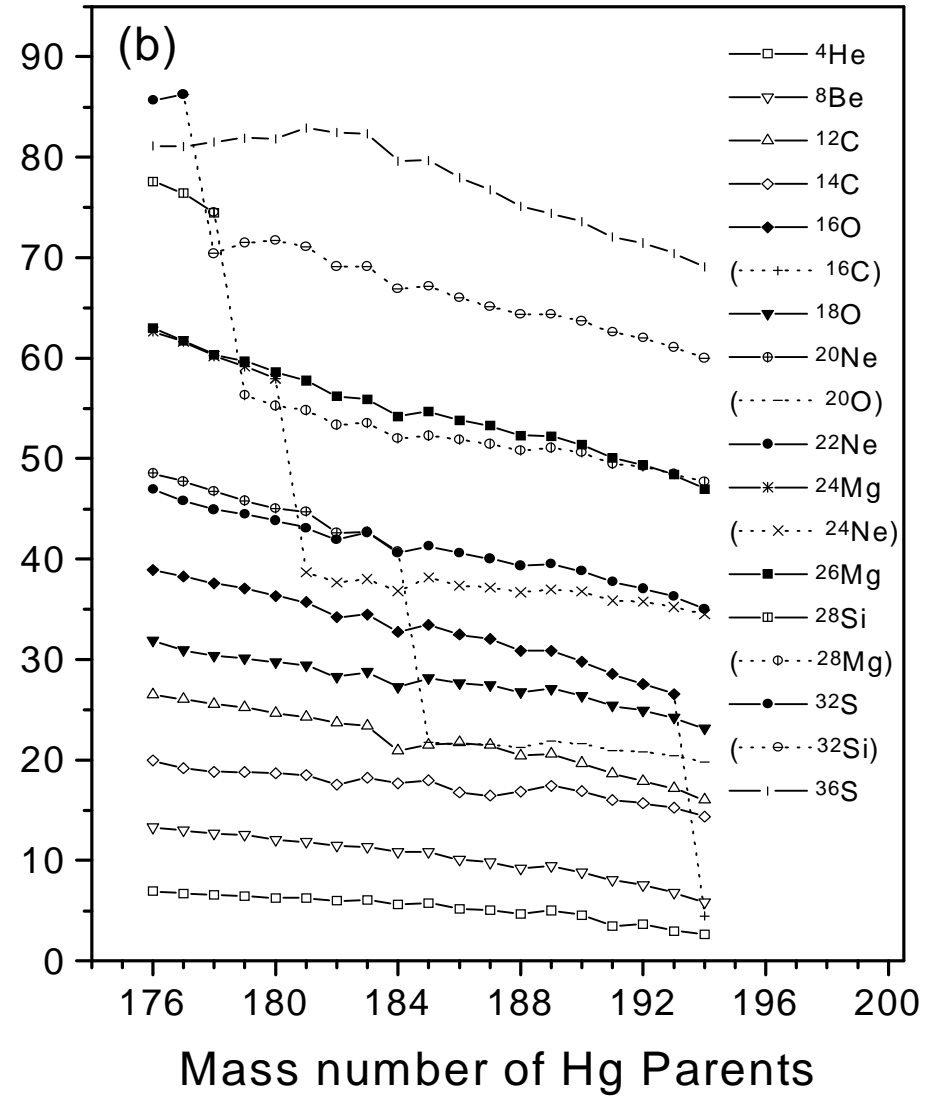
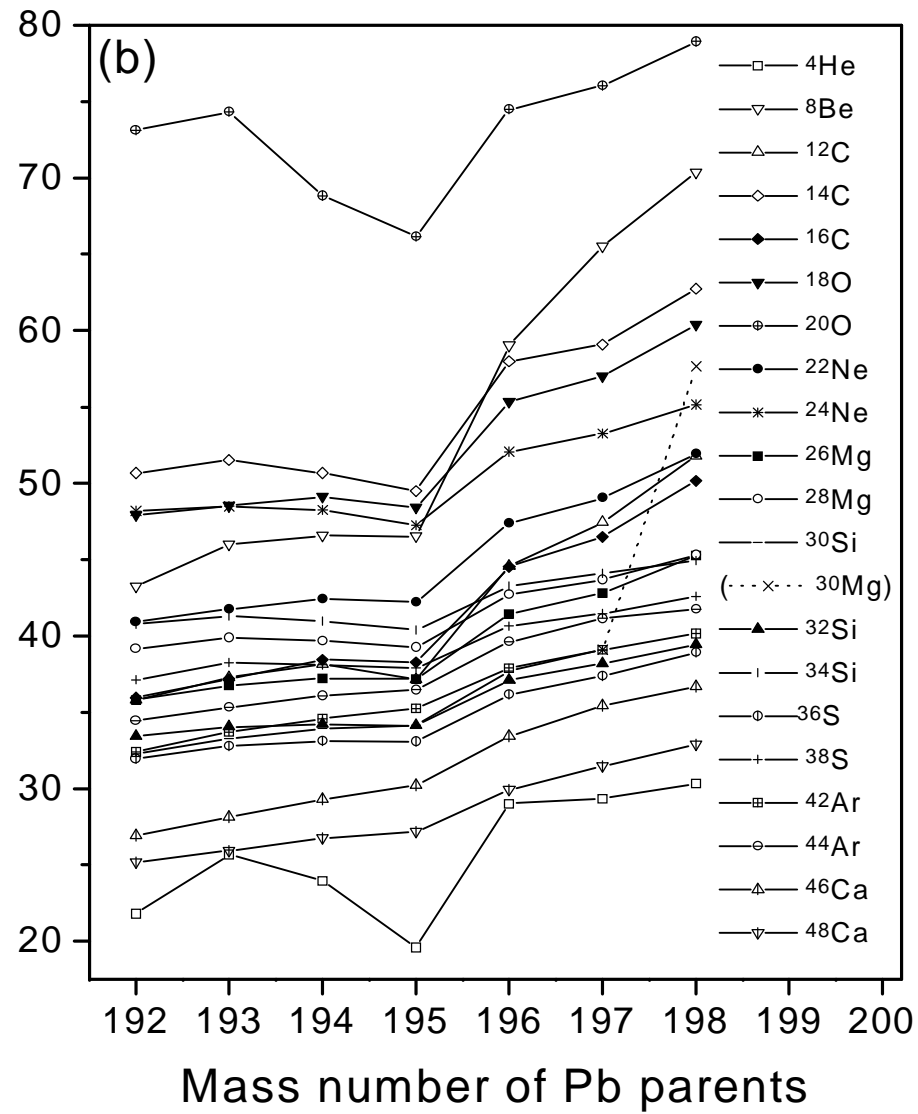
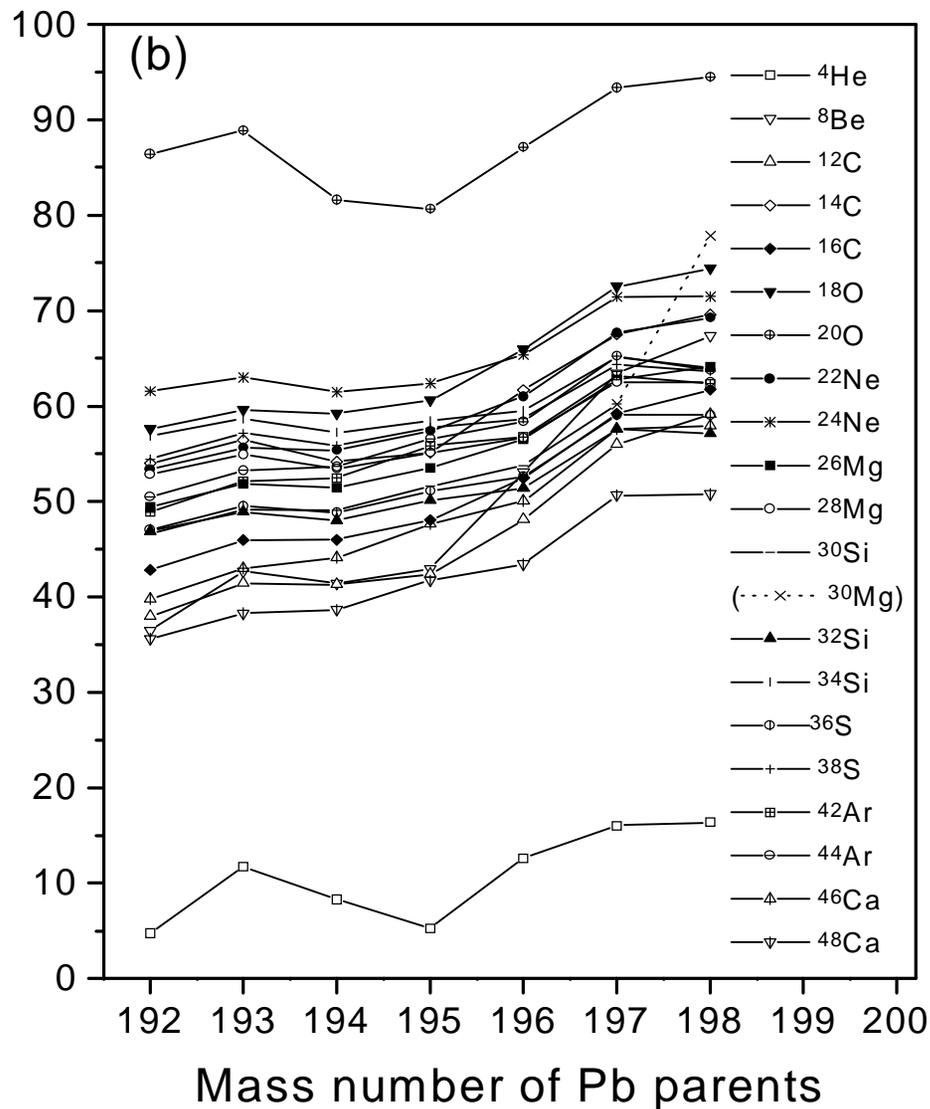
\end{document}